\documentclass{aa}
\usepackage[varg]{txfonts}
\usepackage{multirow}
\usepackage{natbib}
\usepackage{graphicx}
\newcommand{\tabincell}[2]{\begin{tabular}{@{}#1@{}}#2\end{tabular}}

\begin{document}

\title{Possible tidal dissipation in millisecond pulsar binaries}
\author{D. Wang \and B.P. Gong\thanks{bpgong@hust.edu.cn}}
\institute{Department of Physics, Huazhong University of Science and Technology, Wuhan 430074, China}

\abstract
{
    The post-Keplerian(PK) parameters inferred from pulsar timing provide a convenient way to test Einstein's general theory of relativity. However, before obtaining a pure orbital decay $\dot{P}_b$ induced by gravitational wave radiation, which is one of the PK parameters, a number of factors need to be accounted for carefully. The effect of tidal dissipation on $\dot{P}_b$ has been thought of as negligible. Here, we investigate the data for possible effects of tidal dissipation on $\dot{P}_b$.

}
{The possibility of the tidal dissipation as a contributor to $\dot{P}_b$ in a large sample of millisecond pulsar binaries is investigated in detail.
}
{We collected a large sample of pulsar binaries with measured $\dot{P}_b$. All of the systems are millisecond pulsars. The residual $\dot{P}^{Res}_b$ of these systems was obtained by subtracting the three normal effects, that is to say the effect of Shklovskii, line-of-sight acceleration, and gravitational radiation. Assuming that tidal dissipation is responsible for such a residual $\dot{P}^{Res}_b$, the tidal parameters of these systems can be calculated and compared with the tidal models.}
{
    The residual $\dot{P}^{Res}_b$ is distributed over the half positive and half negative. The dynamical tidal model can explain the residual $\dot{P}_b$ of millisecond pulsar-white dwarf binaries. And the Love number of the main-sequence companion of \object{PSR J1227-4853} can be derived as a reasonable value $k_2=0.177^{+0.098}_{-0.058}$ with the equilibrium tidal model. Those results are compatible with the scenario of tidal dissipation. Additionally, a weak correlation between the tidal parameter and orbital period is revealed, likely originating in the tidal process of the recycled stage which is worthy of further investigation.
    }
{}
\maketitle
\section{Introduction}

    Binary pulsars are the natural laboratory of general relativity (GR) in the strong-field regime. High-precision pulsar timing can measure not only the Keplerian orbital parameters, but also the relativistic correction to the Keplerian orbit, which is parameterized to the five post-Keplerian (PK) parameters, that is to say the orbital decay $\dot{P}_b$, the periastron advance $\dot{\omega}$, the Einstein delay $\gamma$, the range $r$, and the shape $s$ of the Shapiro delay \citep[and references therein]{zhang2019constraints}. The most studied parameter, $\dot{P}_b$, has been treated containing only two effects besides GR \citep{zhu2018tests}. One is the Shklovskii kinematic effect originating in the transverse motion of a pulsar binary \citep{shklovskii1970possible}, and the other is the Doppler effect induced by the acceleration along the line of sight of the pulsar binary. After subtracting these effects, the remnant should be the gravitational wave (GW) radiation term originating in GR. However, the remnant $\dot{P}_b$ of some pulsar binaries becomes positive which is inconsistent with the prediction of the GW \citep{pathak2019correcting}, which suggests the involvement of other mechanisms such as wind accretion, tidal dissipation, and magnetic braking. Therefore, the situation becomes complicated. It is thereby worth further investigation to determine the true GW term.

Here, the timing results of all pulsar binaries with the measured derivative of the orbital period are collected, except for binary pulsars with an orbital eccentricity larger than 0.1 and for neutron star-neutron star binaries. Interestingly, all systems are millisecond pulsar binaries. By carefully subtracting the kinematic term, the acceleration term, and the GW term, the residual term of these pulsar binaries is found to be roughly half positive and half negative. Further, the possibility of such a distribution originating from tidal dissipation is investigated.

Tidal dissipation is usually ignored or not discussed in depth in most millisecond pulsar (MSP-) white dwarf (WD) binaries. It has been thought that the companion has already been synchronized with the orbit or that tidal dissipation is too weak on a WD to have any noticeable effects \citep{freire2012relativistic,antoniadis2013massive}. However, all MSP systems have experienced the recycled stage \citep{lorimer2008binary}, so it is still uncertain whether the spin of the companion star has already been synchronized with the orbit after the recycled stage. On the other hand, although the equilibrium tide has been thought to be responsible for the tidal effect on a WD \citep{applegate1994orbital}, which is usually weak, dynamical tidal dissipation on a WD can be more efficient than equilibrium tidal dissipation \citep{fuller2012dynamical}. Hence, it is worth revisiting the possibility of tidal dissipation induced $\dot{P}$ on these systems. 

To evaluate the effect of tidal dissipation on pulsar binaries, the tidal parameters of every system are calculated. Furthermore, we show that the dynamical tide on a WD is efficient enough to produce the maximum tidal parameter in MSP-WD systems. Additionally, the tidal parameter of MSP-MS system \object{PSR J1227-4853} corresponds to a reasonable Love number $k_2=0.177^{+0.098}_{-0.058}$ of a MS companion under the equilibrium tide model. These results likely support the idea that tidal dissipation is also a non-negligible contributor to the $\dot{P}_b$ of MSP binaries, with the equilibrium tidal model responsible for the MS companion and the dynamical tidal model responsible for the WD companion.

Moreover, a weak power-law correlation between the tidal parameter and the orbital period is found in MSP-WD binaries. Compared with the general tidal model, such a correlation requires that the spin of the companion correlates with the orbital frequency in millisecond pulsar binaries, which possibly results from binary evolution.

This paper is organized as follows: The estimation of the residual $\dot{P}_b^{Res}$  in all systems is described in  section \ref{section data}. The comparison of the tidal parameters of these systems with tidal models is presented in  section \ref{section compare}. The discussion and conclusion are drawn in  section \ref{section discussion} and \ref{section conclusion}.

\section{Samples collection and $\dot{P}_b$ calculations}
\label{section data}
\subsection{The general components of observational $\dot{P}_b$}
The observed $\dot{P}_b$ from pulsar timing is the sum of different effects. It is contaminated by two kinematic effects, with one being the Shklovskii kinematic effect originating in the transverse motion of a pulsar binary \citep{shklovskii1970possible}:
\begin{equation}\label{Shk}
    \dot{P}_b^{Shk}=\mu_{\perp}^2 \frac { D } { c } P _ { b }.
\end{equation}
Here $\mu_{\perp}$ is the total proper motion velocity, $P_b$ is the orbital period, $D$ is the distance of the binary, and $c$ is the speed of light. The other effect contributing to the measured $P_b$ stems from the acceleration of the pulsar along the line of sight:
\begin{equation}\label{acc}
    \dot{P}_b^{Acc}=\frac{A}{\mathbf{c}}P_{b},
\end{equation}
where $A$ is the acceleration along the line of sight. In subtracting these two terms as shown in equation \ref{Shk} and equation \ref{acc} from the observed one, we obtain the intrinsic $\dot{P}_b^{Int}$, which is considered to be dominated by gravitational radiation so that $\dot{P}_b^{Int}=\dot{P}_b^{GW}$, where
\begin{equation}\label{gr}
    \dot{P}_{b}^{GW}=-\frac{192 \pi P_{b}^{-5 / 3} G^{5 / 3} m^{4 / 3} m_{p} m_{c}}{5(2 \pi)^{-5 / 3} c^{5}}\left(1+\frac{73}{24} e^{2}+\frac{37}{96} e^{4}\right)\left(1-e^{2}\right)^{-7 / 2}.
\end{equation}
Here $G$ is the gravitational constant; $m$ is the total mass of binary; $m_p$ and $m_c$ are the mass of the pulsar and companion star, respectively; and $e$ is the orbital eccentricity.

\subsection{Pulsar binary data}

To test the effect on $\dot{P}_b$ other than $\dot{P}_b^{Shk}$, $\dot{P}_b^{Acc}$, and $\dot{P}_b^{GW}$ mentioned above, all pulsar binaries with $\dot{P}_b$ detected within a $3\sigma$ significance and of a relatively low eccentricity ($e<0.1$) were collected, excluding the double neutron star system. There are 26 pulsar binaries available and all of them are millisecond pulsar binaries. The measured parameters of those sources are shown in Table \ref{tab:data}. A detailed discussion of these parameters is given in the following.
 
Among the 26 sources, two sources are the main-sequence star (MS) millisecond pulsar binaries; 17 are PSR-WD systems; and the remaining seven sources have no optical counterpart or the type of their companion star cannot be determined by the optical observation, which suggests that the companions of these pulsars are faint. Those uncertain companions are thus assumed to be WDs.

\subsection{The estimation of the kinematic term in $\dot{P}_b$}\label{sec:pb_k}
The estimation of $\dot{P}_b^{Shk} $ and $ \dot{P}_b^{Acc}$ from equation \ref{Shk} and \ref{acc} requires the distance, proper motion, and acceleration of the systems. In our samples, almost all binaries have a measured proper motion, except \object{PSR J1701-3006B} and \object{PSR J1807-2459A}. The proper motion of \object{PSR J2222-0137} is measured by the very long baseline interferometry (VLBI;\citealt{deller2013vlbi}), the proper motion of \object{PSR J1227-4853} and \object{PSR J1957+2516} are measured by Gaia \citep{gaia2018vizier}, and the proper motion of the remaining sources are given by pulsar timing.
 
On the distance measurement of the 26 sources, two pulsars have a precise distance measurement, one being \object{PSR J2222-0137} which was measured by VLBI \citep{deller2013vlbi} and the other one being \object{PSR J1227-4853} which was measured by Gaia gaia2018vizier. For the remaining 24 sources, the distance of ten pulsar binaries is given by the pulsar timing via a parallax, and the remaining sources have no distance measurement. As 11 pulsars are located in the globular clusters, their distance can be estimated by the distance of the globular clusters given by Baumgardt \citep{baumgardt2018mean}. For the remaining three pulsars with the absence of all those methods for the distance measurement mentioned above, their distance was estimated by their dispersion measure (DM) and YMW16 galactic electron density model \citep{yao2017new}, which can be found by the "$DIST\_ DM$" parameter from the Australia Telescope National Facility (ATNF) pulsar catalog \citep{manchester2005australia}. For these systems, a 1 $\sigma$ uncertainty is assumed to be 20 $\%$ of the corresponding distance value. 
 
The acceleration of a pulsar is normally attributed to the galactic potential, which can be calculated by McMillan's model \citep{mcmillan2016mass}. For binaries in the globular clusters, the potential of the cluster also contributes to the acceleration of the pulsar binary, which can be calculated by the simple globular cluster model of Freire \citep{freire2005millisecond}. 

In Freire's model, the potential of the globular cluster can be expressed by the density and radius of the core of the cluster, which has been estimated by Baumgardt for all globular clusters \citep{baumgardt2018mean}. Then the acceleration along the line of sight only depends on $z$, the distance between the point and the sky plane through the center of the cluster, which cannot be measured. By searching the largest possible range, the acceleration has a maximum $A_{max}^{gc}$ and a minimum $A_{min}^{gc}$, which has the same magnitude but an opposite direction. We assume that the probability distribution of the line-of-sight acceleration in globular clusters is described by the Gaussian distribution with a mean value $0$ and a standard deviation of $A_{max}^{gc}/3$. 

Now, the contribution of the Shklovskii and acceleration effect to $\dot{P}^{int}_b$ can be calculated by equation \ref{Shk} and \ref{acc}, respectively. After subtracting two kinematic effects from the measured one, $\dot{P}_b^{Obs}$, the intrinsic orbital decay, $\dot{P}_b^{Int}$, can be obtained,
\begin{equation}\label{int}
    \dot{P}_b^{Int}=\dot{P}_b^{Obs}-\dot{P}_b^{Shk}-\dot{P}_b^{Acc}.
\end{equation}
Such an intrinsic $\dot{P}^{int}_b$ should be dominated by the gravitational radiation term. However, a considerable number of systems show positive  $\dot{P}^{int}_b$, which is inconsistent with the prediction of GW radiation, expecting a negative $\dot{P}_b$ only. Most probably,  $\dot{P}^{int}_b$ is contaminated by an effect other than the two terms at the right-hand side of the above equation.

\subsection{The residual $\dot{P}_b^{Res}$}

To figure out such an effect, we need to subtract the gravitational radiation-induced orbital changing $\dot{P}_b^{GW}$ from the intrinsic one, which is $\dot{P}_b^{Res}=\dot{P}_b^{Int}-\dot{P}_b^{GW}$. Such a residual $\dot{P}_b^{Res}$ reflects the behavior of the unknown effect, which can be examined in terms of significance. 

To calculate the term of gravitational radiation given by equation \ref{gr}, the orbital eccentricity and masses of double stars are necessary. The eccentricity can be measured by pulsar timing. For binaries without an eccentricity measurement, the nondetection means that the eccentricity is very small and can be treated as zero.

The companion mass, $m_c$,  pulsar mass, $m_p$, and  orbital inclination, $i$, are  related to the mass function $f(m_c)$ by
\begin{equation}\label{massfunction}
    f(m_c)=\frac{m_c^3 \sin^3 i}{(m_p+m_c)^2}.
\end{equation}

Such a mass function can be measured by pulsar timing for all sources. The orbital inclination and companion mass can be given by the measurement of the Shapiro delay. However, the Shapiro delay has only been detected in five binaries in our samples. And there are four binaries with mass ratios measured by their radial velocities (See Table \ref{shaprio}).

To estimate the mass of double stars, three assumptions are made. For all binaries without a Shapiro delay detection, the pulsar mass is assumed to be $1.48\pm0.2M_{\odot}$, given the statistics result \citep{ozel2012mass}. For the binaries with neither a Shapiro delay detection nor a mass ratio measurement, the direction of the orbit is assumed to be isotropic, corresponding to a probability density function of orbital inclination $P(i)=\sin i$. And the minimum mass of the companion of all binaries is assumed to be $0.08 M_{\odot}$.

Thus, the gravitational radiation-induced $\dot{P}_b^{GW}$ can be obtained by equation \ref{gr}. After subtracting it from the intrinsic one, $\dot{P}_b^{Int}$, we determined the residual term, $\dot{P}_b^{Res}$. All components of $\dot{P}_b$ from our sample are shown in Table \ref{tab:data}. 

We did not use the standard technique of statistical error propagation because the non-Gaussian probability density function was adopted in the estimation for the mass and the acceleration was calculated by the code from McMillan's model \citep{mcmillan2016mass}, which does not include error propagation. Thus, the uncertainty of $\dot{P}_b^{Res}$ was calculated by the Monte Carlo method. The middle value was estimated by the 50th percentile of the sampling result, and the 1 $\sigma$ lower and  upper uncertainty was estimated by the 17.37th and 82.63th percentile of the sampling result, respectively. The same method was applied to the uncertainty estimation of the following calculations.

\begin{figure*}
    \centering
    \includegraphics[width=\textwidth]{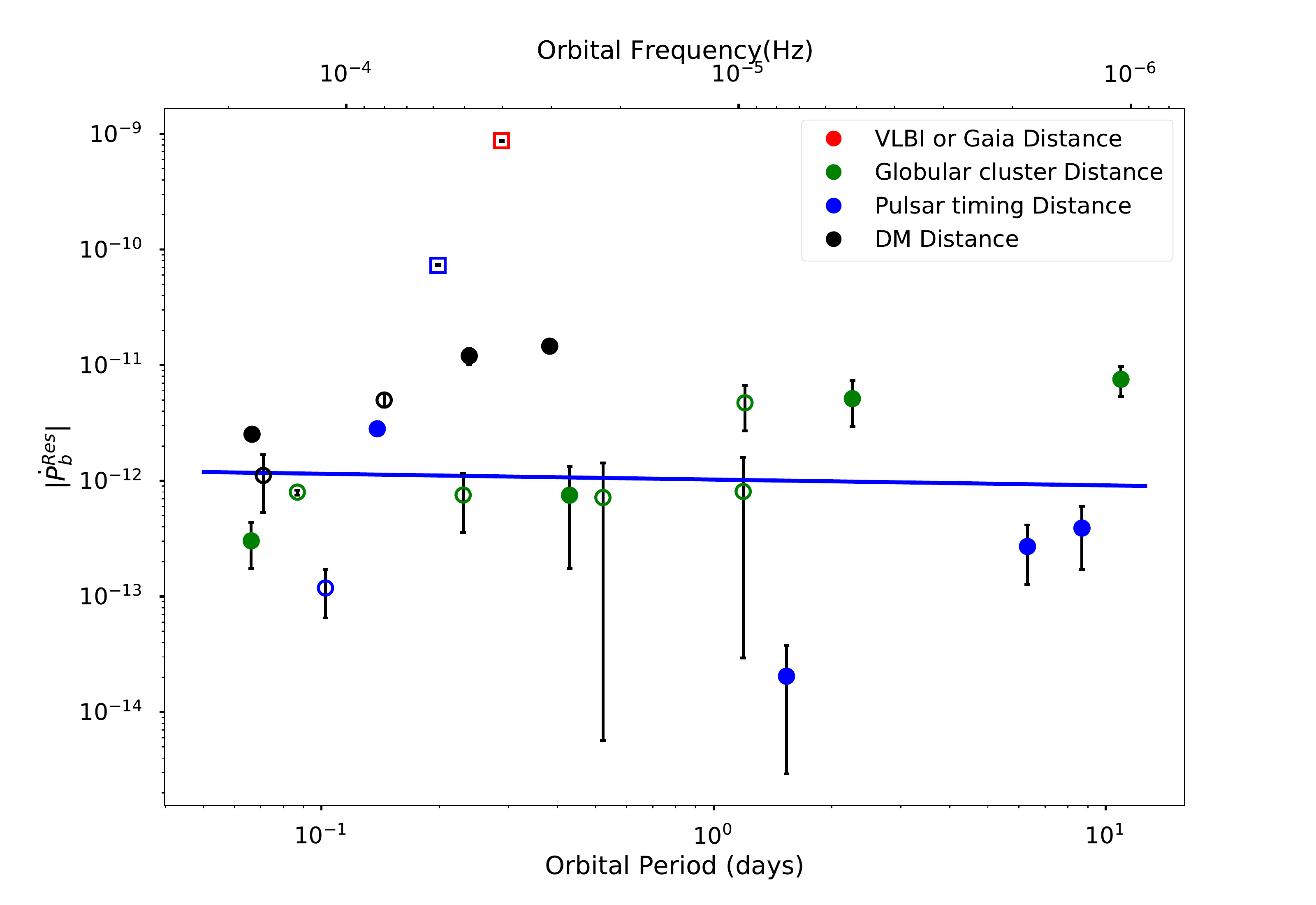}
    \caption{Relationship between $|\dot{P}_b^{Res}|$ and orbital period $P_b$ in logarithmic coordinates. Only the systems with $\dot{P}_b^{Res}$ in 1 $\sigma$ significance are shown. Different colors represent the binaries of the different ways of the distance measurements\protect\footnotemark[1]. The solid symbols denote systems with positive $\dot{P}_b^{Res}$ and the hollow ones represent sources with negative $\dot{P}_b^{Res}$. The line is the fitting by the power law relation of systems with a WD\ companion. \label{pbdotfig}}
\end{figure*}

The significance of the residual $\dot{P}_b^{Res}$ is within 1 $\sigma$ for 21 binaries, 2 $\sigma$ for 14 binaries, and 3 $\sigma$ for nine binaries as shown in Fig \ref{pbdotfig}. The uncertainty of $\dot{P}_b^{Res}$ mainly stems from the measurement for the distance. As shown in Table \ref{tab:data}, for the majority of systems, $\dot{P}_b^{Obs}$ is close to their $\dot{P}_b^{Shk}$. The uncertainty of the distance measurement by pulsar timing is usually large, corresponding to a large uncertainty on $\dot{P}_b^{Res}$.

Such a residual $\dot{P}_b^{Res}$ reflects the behavior of the unknown effect. Among 21 binaries, eleven sources have positive values for $\dot{P}_b^{Res}$ and ten sources have negative ones. The half-to-half distribution of $\dot{P}_b^{Res}$ is easy to understand in the scenario for random noise, whereas the possibility of the tidal effect on the residual $\dot{P}_b^{Res}$ discussed in the following section provides another scenario that may lead to such a distribution.

\footnotetext[1]{Here, two systems with a distance measured by the globular cluster that does not have a proper motion measurement belong to the "DM Distance" group}
\section{The comparison with the tidal model}\label{section compare}
\subsection{Tidal models}
Considering a circular binary containing a companion star with radius $R_c$ and mass $m_c$ and a primary star with mass $m_p$ treated as a point mass, the companion star is deformed by the tidal potential of the primary star, then the deformation of the companion star produces an extra torque and disturbs the motion of the binary. The first derivative of the orbital period due to tidal torque can be expressed as
\begin{equation}\label{dotpb}
    \dot{P}_b=\frac{3T P_b^2}{2 \pi \mu a^2}.
\end{equation}
Here $\mu=m_p m_c/(m_p+m_c)$ is the reduced mass, $a$ is the orbital semi-major axis, and $T$ is the tidal torque, which can be parameterized as \citep{fuller2012dynamical}
\begin{equation}\label{T0}
    T=T_0 F(\omega)\mathrm{sgn}(\omega).
\end{equation}
Here $T_0$ is $G(m_p/a^3)^2R_c^5$, representing the effect of the external field, and $F(\omega)$ is the response of the companion star to the gravity exerted by the primary star, which is usually denoted as a dimensionless function of tidal frequency $\mathbf{\omega=\Omega_c-\Omega}$ (defined as the difference between the orbital frequency $\Omega$ and spin frequency$\Omega_c$), and the direction of torque only depends on the \text{sign} of the tidal frequency.

For the general binaries, the formula for $F(\omega)$ also depends on the tidal frequency, the eccentricity, the inclination between the spin of the companion star and orbit, and the interior physics of the companion star \citep{ogilvie2014tidal}. If the orbit is circular and if the inclination is small, the eccentric and oblique terms in the tidal response vanish, thus $F(\omega)$ only depends on the tidal frequency and interior physics of the star.

There are two kinds of tidal models \citep{ogilvie2014tidal}. One kind is the equilibrium tide in which a star is in hydrostatic equilibrium, and the other is the dynamical tide in which a star is treated as an oscillator in various modes. 

The tidal dissipation of equilibrium tide is normally attributed to the turbulent viscosity in the convection zone, which is the most powerful mechanism of energy dissipation \citep{zahn1989tidal}. In such a model, the dimensionless tidal response is inversely proportional to the convection friction time $t_f=(MR^2/L)^{1/3}$ and proportional to the tidal frequency
\begin{equation}\label{zahn}
    F(\omega) =3\frac{k_2}{t_f}\frac{\omega}{\Omega_{k}^2}\sqrt{\frac{320}{320+\eta^2}}
.\end{equation}
Here $\Omega_{k}=(Gm_c/R^3)^{\frac{1}{2}}$ is THE dynamical frequency of the companion star, coefficient $k_2$ is THE Love number of the companion star, and $\eta=t_f\omega/\pi$ \citep{zahn1989tidal,goupil2008tidal}.

Even though the Zahn's equilibrium tide model is for the MS, it had been widely applied to the investigation of WD tidal dissipation \citep{freire2012relativistic,antoniadis2013massive}. The more sophisticated model, that is to say the dynamical model for the WD had been discussed by Fuller \citep{fuller2012dynamical,fuller2013dynamical}, and it can produce more efficient tidal dissipation, especially at a high tidal frequency.

In this model, the excitation and dissipation of excited gravity waves in the star are involved, and the outgoing propagating waves can be efficiently damped in the outer layer of the star by the nonlinear mechanism. Fuller numerically simulated the dynamical tide on the WD and found the tidal response $F(\omega)$ satisfied
\begin{equation}
    F(\omega)\propto \omega^5
\end{equation}
for a carbon-oxygen WD, and
\begin{equation}
    F(\omega)\propto \omega^6
\end{equation}
for a helium WD.

If the residual $\dot{P}_b$ originated in the tide dissipation, the magnitude of the tidal response $F(\omega)$ should be consistent with the tidal model. In such a case, $\dot{P}_b^{Res}$ can also set a constraint on the interior of the companion on pulsar binaries.

\subsection{On the value of $F(\omega)$}
By combining equations \ref{dotpb} and \ref{T0}, the tidal response $F(\omega)$ can be calculated as
\begin{equation}
    F(\omega)=\frac{\mu a^\mathbf{8}\Omega^2}{6\pi G m_p^2 R_c^5 }|\dot{P_b^{Res}}|=\frac{\mu G^{5/3} m^{8/3}\Omega^{-10/3}}{6\pi m_p^2 R_c^5}|\dot{P_b^{Res}}|,\end{equation}
where $\Omega=(2\pi/P_b)$ is the orbital frequency.

The calculation of $F(\omega)$ of all systems from equation \ref{dotpb} requires knowledge on the radius of the companion, which can be obtained  by  the  analytic relation between the mass and radius of WDs \citep{andronov1990moments}:
\begin{equation}\label{wdr}
    \frac{R_c}{R_{\odot}}=0.01153\left(\left(\frac{m_c}{1.44 M_{\odot}}\right)^{-\frac{2}{3}}-\left(\frac{m_c}{1.44 M_{\odot}}\right)^{\frac{2}{3}} \right)^{0.465}.
\end{equation}
For the MS companion, we used the empirical formula of the mass-radius relation \citep{demircan1991stellar}:
\begin{equation}\label{msr}
    \frac{R_c}{R_{\odot}}=1.06\left(\frac{m_c}{M_{\odot}}\right)^{0.945}.
\end{equation}

In order to check whether the $F(\omega)$ of systems are expected by the tidal model or not, tidal frequencies are required. However, they are unknown for all PSR-WD binaries because the rotation velocity of those companions is not measurable. Therefore, the tidal model can only impose an upper limitation on $F(\omega)$ for those systems.

The maximum $F(\omega)$ among all the systems is about $10^{3}$, corresponding to a convection friction time $t_f\approx6 \times 10^{-12}\text{year}$ in Zahn's equilibrium model. Such a friction time corresponds to an impossibly large luminosity of $L=GM^2/t_f^3\approx 10^{28} L_{\odot}$. Therefore, Zahn's equilibrium tide dissipation is too weak to explain the $F(\omega)$ of the pulsar binaries.

In contrast, in Fuller's dynamical tide model, the maximum magnitude of $F(\omega)$ is about $40$ for helium WD and $10^5$ for carbon-oxygen WD in the case that companion rotates with break-up velocity. Such an upper limit for $F(\omega)$ is larger than the maximum one of all the systems. Therefore, the dynamical tide can provide a sufficient tidal response for these systems.

If the spin of the companion can be measured, the $F(\omega)$ can be obtained and compared with the tidal model. \object{PSR J1227-4853} is a millisecond pulsar binary with a MS companion, the projected rotation velocity of which was measured through spectrum broadening $v_c=86\pm20\text{~km}\text{~s}^{-1}$\citep{de2014unveiling}. The spin of the companion can thus be obtained by $\Omega_c=v_c/R_c\sin i=41^{+10}_{-9}\text{~rad/day}$, and the inclination $i$ and radius $R_c$ was obtained by the same method as in section \ref{sec:pb_k}.

The convection friction time of the companion of \object{PSR J1227-4853} in Zahn's tidal model is 
\begin{equation}
    t_f=(MR^2/L)^{1/3}= 0.26 \pm 0.01 \textrm{~year},
\end{equation}
where the luminosity was obtained by the temperature $6100 \text{~K }$\citep{de2014unveiling}, assuming the radiation is a blackbody. If $\dot{P}_b^{Res}$ is caused by the tidal dissipation of Zahn's equilibrium tide model, it implies a tidal response of $F(\omega)=1.3^{+0.67}_{-0.41}\times 10^{-7}$ , which yields a Love number for the companion star of $k_2=0.177^{+0.098}_{-0.058}$. Such a Love number is a reasonable value for the MS. Consequently, the $\dot{P}_b^{Res}$ of the MS-MSP binaries can be explained by Zahn's model of equilibrium tide.

The tidal model can not be tested directly due to the lack of companion rotation velocity in the PSR-WD systems. As the maximum $F(\omega)$ among these PSR-WD binaries can be achieved by the dynamical tidal model with a reasonable rotation velocity, the contribution of the tidal dissipation to the orbital decay $\dot{P}_b$ of the systems cannot be excluded yet. On the other hand, a reasonable Love number for the MS companion of \object{PSR J1227-4853} inferred by equation \ref{zahn} can be regarded as supporting evidence for the equilibrium tide on the PSR-MS systems.

\subsection{ A possible correlation}

We also find a possible correlation between the tidal response parameter $F(\omega)$ and the orbital period $P_b$. To analyze the tidal dissipation of these systems, a new tidal parameter $K$ rather than $F(\omega)$ was adopted:
\begin{equation}\label{K}
K=F(\omega)\Omega^{10/3}=\frac{\mu G^{5/3} m^{8/3}}{6\pi m_p^2 R_c^5}|\dot{P}_b^{Res}|.
\end{equation}
As $F(\omega)$ is dependent on $P_b$, the correlation analysis of tidal parameters and the orbital period is affected. In contrast,  the calculation of $K$ only uses $\dot{P}_b^{Res}$ and the mass of a binary system, which is more suitable for the tidal analysis.

\begin{figure*}
    \centering
    \includegraphics[width=\textwidth]{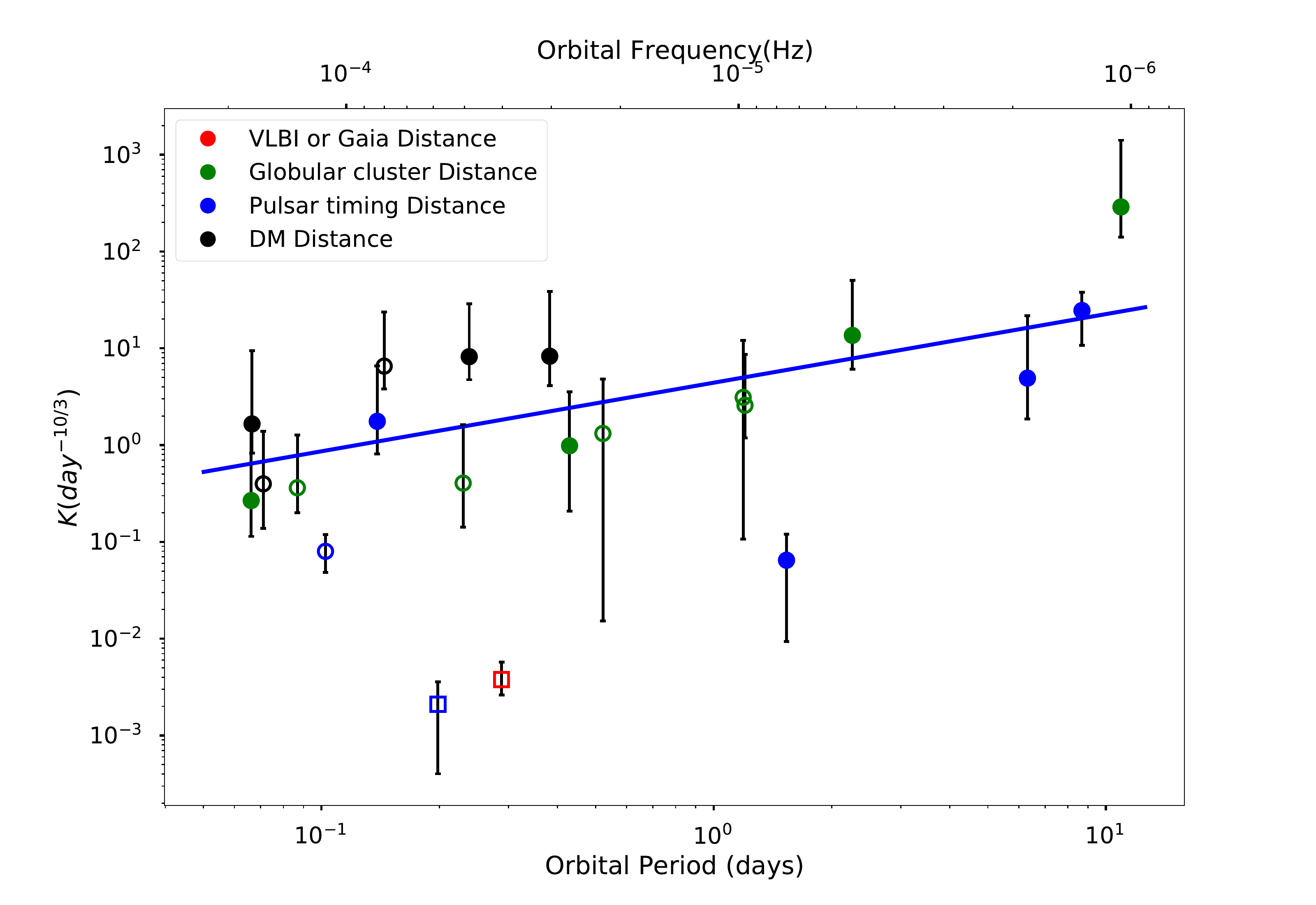}
    \caption{Relationship between $K$ and orbital period $P_b$ in logarithmic coordinates. Only systems with $\dot{P}_b^{Res}$ above a 1 $\sigma$ significance are shown. The meaning of the point style is the same as in Fig \ref{pbdotfig}. \label{kfig}}
\end{figure*}

Upon obtaining the tidal parameter $K$,  a weak positive correlation emerged as shown in Fig. \ref{kfig}. Thus the Pearson correlation coefficient between $\log P_b$ and $\log K$ of MSP-WD systems can be obtained and it exhibits a quantitative evaluation of such a correlation. As different methods of the distance measurement affect the reliability of the result, the Pearson correlation coefficient of $\log|\dot{P}_b^{Res}|$ and $\log P_b $  is shown in table \ref{tab:Pearson}, which is apparently influenced by the different ways used to measure the distance.

\begin{table*}
    \centering
    \caption{Results of the corelation analysis. The Pearson correlation coefficient of our systems correspond to different combinations of  the distance measurement. The different groups are denoted as follows: "A" corresponds to sources with a distance measured by the globular cluster; "B" is for sources with a distance  measured by pulsar timing; and "C" corresponds to sources with a distance  measured by a DM, and two systems with a distance measured by the globular cluster that do not have a proper motion measurement also belong to Group C.\label{tab:Pearson}}
    \begin{tabular}{ccccccc}
        Significance&Groups  &  Num  &\tabincell{c}{Pearson coefficient\\for $\log K-\log P_b$}&\tabincell{c}{p-value\\for $\log K-\log P_b$}&\tabincell{c}{Pearson coefficient\\for $\log|\dot{P}_b^{Res}|-\log P_b$}&\tabincell{c}{p-value\\for $\log |\dot{P}_b^{Res}|-\log P_b$}\\
        \hline
        \multirow{3}{*}{$1\sigma$} 
        &A&9&0.95&$7.37\times 10^{-5}$&0.86&$3.26\times 10^{-3}$\\
        &A,B&14&0.76&$1.73\times 10^{-3}$&0.17&$5.43\times 10^{-1}$\\
        &A,B,C&19&0.62&$5.56\times 10^{-3}$&-0.03&$8.82\times 10^{-1}$\\
        \hline
        \multirow{3}{*}{$2\sigma$} 
        &A&5&0.97&$6.94\times 10^{-3}$&0.95&$1.34\times 10^{-2}$\\
        &A,B&7&0.93&$2.00\times 10^{-3}$&0.81&$2.73\times 10^{-2}$\\
        &A,B,C&11&0.82&$2.16\times 10^{-3}$&0.54&$8.64\times 10^{-2}$\\
        \hline
        $3\sigma$&A,B,C&6&0.97&$1.06\times 10^{-3}$&0.44&$3.85\times 10^{-1}$\\
        \hline
\end{tabular}
\end{table*} 
\begin{table*}
    \centering
    \caption{Results of the corelation analysis. The groups are same as Table \ref{tab:Pearson}, but not including \object{PSR J0024-7204X} (i.e., the source in the top right corner of Figure \ref{kfig}). \label{tab:Pearson2}}
    \begin{tabular}{ccccccc}
        Significance&Groups  &  Num  &\tabincell{c}{Pearson coefficient\\for $\log K-\log P_b$}&\tabincell{c}{p-value\\for $\log K-\log P_b$}&\tabincell{c}{Pearson coefficient\\for $\log|\dot{P}_b^{Res}|-\log P_b$}&\tabincell{c}{p-value\\for $\log |\dot{P}_b^{Res}|-\log P_b$}\\
        \hline
        \multirow{3}{*}{$1\sigma$} 
        &A&8&0.95&$3.00\times 10^{-4}$&0.78&$2.39\times 10^{-2}$\\
        &A,B&13&0.68&$9.87\times 10^{-3}$&-0.02&$9.58\times 10^{-1}$\\
        &A,B,C&18&0.49&$3.75\times 10^{-2}$&-0.18&$4.69\times 10^{-1}$\\
        \hline
        \multirow{3}{*}{$2\sigma$} 
        &A&4&0.97&$2.76\times 10^{-2}$&0.97&$3.25\times 10^{-2}$\\
        &A,B&6&0.86&$2.81\times 10^{-2}$&0.78&$6.58\times 10^{-2}$\\
        &A,B,C&10&0.64&$4.61\times 10^{-2}$&0.55&$1.00\times 10^{-1}$\\
        \hline
        $3\sigma$&A,B,C&5&0.81&$9.54\times 10^{-2}$&0.92&$2.82\times 10^{-2}$\\
        \hline
\end{tabular}
\end{table*} 

As shown in Table \ref{tab:Pearson}, the Pearson correlation coefficients of $\log K$ and $\log P_b$ imply that there is a weak correlation between them in different combinations of the distance measurement and significance levels. Such a correlation likely does not originate in the correlation between $\log |\dot{P}_b^{Res}|$ and $\log P_b$ due to their small Pearson correlation coefficients. As the correlation is strongly affected by the point at the top right of the Figure \ref{kfig}, we carried out the same analysis without that point as shown in Table \ref{tab:Pearson2}. In such a case, the only difference from the previous analysis is that the correlation between $\dot{P}_b^{Res}$ and $P_b$ in the $3\sigma$ significance exhibits a significant enhancement. Nevertheless, the Pearson correlation coefficients of $\log K$ and $\log P_b$ still show a weak correlation between them, as shown in the previous case.

A strong correlation between $\dot{P}_b^{Res}$ and $P_b$ is also shown in the pulsar binaries of the s. Such a systematic correlation probably stems from the assumption that the acceleration of the globular clusters satisfies the Gaussian distribution with a mean value $0$ and a standard deviation $a_{max}^{gc}/3$. As the correlation coefficient was calculated by the middle value of the data,   $\dot{P}_b^{Acc}$, which is proportional to the orbital period, may not have been correctly estimated in calculations of the Pearson correlation coefficient. Nevertheless, the strong correlation favors the existence of universal tidal dissipation in the majority of MSP-WD systems.

Fitting with a simple power-law relation $K=\alpha P_b^{\beta}$ was carried out for MSP-WD systems, as shown in Table \ref{tab:fitting}. The probability density distribution of $K$ and $\dot{P}_b^{Res}$ is non-Gaussian, which was thus analyzed by the  Monte Carlo method. The steps are as follows: the values of $K$ or $\dot{P}_b^{Res}$ were sampled on the basis of their probability density distribution, and a classical linear regression in logarithmic coordinates for those values was performed. After repeating such a process 100000 times, the distribution of the slope and intercept was obtained, which exhibits a Gaussian distribution. The reduced Chi-square was calculated by taking the mean of the upper and lower uncertainty at a 1 $\sigma$ level.

\begin{table*}
    \centering
    \caption{Fitting result of $K-P_b$ corresponds to different combinations of  the distance measurement. The different groups are denoted as
follows: "A" corresponds to sources with a distance measured by the globular cluster; "B" is for sources with a distance  measured by pulsar timing; and "C" corresponds to sources with a distance  measured by the DM, and two systems with a distance measured by the globular cluster that do not have a not proper motion measurement also belong to Group C.\label{tab:fitting}}
    \begin{tabular}{cccccc}
        Significance&Groups  &  Num  &  $\beta$  &  $\log\alpha$  &  $\chi^2_{red}$ \\
        \hline
        \multirow{3}{*}{$1\sigma$} 
        &A&9&$1.28^{+0.30}_{-0.29}$&$0.77^{+0.26}_{-0.20}$&0.39\\
        &A,B&14&$0.97^{+0.20}_{-0.20}$&$0.45^{+0.19}_{-0.16}$&2.33\\
        &A,B,C&19&$0.71^{+0.19}_{-0.18}$&$0.64^{+0.17}_{-0.14}$&3.51\\
        \hline
        \multirow{3}{*}{$2\sigma$} 
        &A&5&$1.26^{+0.29}_{-0.30}$&$0.92^{+0.33}_{-0.23}$&0.82\\
        &A,B&7&$1.29^{+0.29}_{-0.25}$&$0.90^{+0.30}_{-0.20}$&3.36\\
        &A,B,C&11&$1.08^{+0.28}_{-0.23}$&$1.15^{+0.27}_{-0.18}$&5.20\\
        \hline
        $3\sigma$&A,B,C&6&$0.96^{+0.34}_{-0.25}$&$1.54^{+0.35}_{-0.21}$&0.36\\
        \hline
\end{tabular}
\end{table*} 

Such a correlation between $K$ and $P_b$ indicates that $\dot{P}_b^{Res}$ originates in tidal dissipation in the majority of MSP-WD systems. For the MSP-MS systems, only two systems are available, and the origin of $\dot{P}_b^{Res}$ for them cannot be judged by this correlation. The possible origin of this correlation is further discussed in section \ref{section discussion correlation}

\section{Discussion}\label{section discussion}
\subsection{The effect of the winds}
The change in system mass can also affect the orbital period. For the circular binary, the derivative of the orbital period due to the change of mass is \citep{tout1991wind}
    \begin{align}\label{mdotwindsacc}
    \dot{P}_b^{\dot{m}}&=-2\frac{\dot{m}}{m}P_b+\frac{3\dot{m}_p(m_p-m_c)}{m_pm_c}P_b\notag\\
    &=-2\frac{\dot{m}_c}{m}P_b+\frac{\dot{m}_p(3mm_p-3mm_c-2m_pm_c)}{mm_pm_c}P_b,
    \end{align}
where $\dot{m}=\dot{m}_c+\dot{m}_p$ is total mass loss.

All systems of our samples are detached binaries, so the transfer and loss of mass can only possibly come from the winds' accretion. In the simple case that the pulsar does not accrete winds ($\dot{m}_p=0$), the derivative of the orbital period is
\begin{equation}\label{mdotwinds}
    \dot{P}_b^{\dot{m}}=-2\frac{\dot{m}_c}{m}P_b.
\end{equation}
Such a $\dot{P}_b^{\dot{m}}$ is always positive because $\dot{m}_c<0$.

The material in the atmosphere of the star can escape from the surface driven by the radiation and produce winds. The existence of stellar wind is often seen in MS. For the WD, the winds only exist in the hot WD \citep{unglaub2008mass}. In all WD-MSP binaries with a measured companion temperature, \object{PSR J0024-7204U} has the highest companion temperature of about $12000 \mathrm{~K}$ \citep{cadelano2015optical}. So all WDs that have a measured temperature in our samples are not hot and cannot produce winds. For these WDs for which temperature has not been measured, the mass of winds cannot be estimated. 

On the other hand, the winds can also be driven by the radiation from the pulsar \citep{freire2012relativistic}. The mass of such winds depends on the power of the radiation from the pulsar, which can be estimated by the intrinsic spin-down. By assuming the total energy release of the pulsar that is received by the companion, the upper limit of the mass of the winds can be estimated. The intrinsic spin-down was calculated from the observational one after subtracting the kinematic effects. This complicates the analysis.

We adopted a simpler method to estimate the mass of the winds via the modulation of the DM \citep{freire2012relativistic}. The winds can produce an extra column density along the line of sight, and thus disturb the observational DM. This method does not depend on the companion type or the origin of the radiation driving the winds.

By assuming the isotropic winds of the WD, the mass loss was obtained by
\begin{equation}\label{mdotofwinds}
    \dot{m}_c=-2\pi X \mu_p v_0 a \frac{\sin i}{i}\Delta H
.\end{equation}
Here $v_0$ is the velocity of the wind, $X$ is the number of nucleons per electron, $\mu_{\mathrm{P}}$ is the proton mass, and $\Delta H$ is the amplitude of the column density modulation which can be related to the DM modulation by $\Delta H=\Delta \mathrm{DM} \times 3.0857 \times 10^{18} \mathrm{~cm} \,\mathrm{pc}^{-1}$.

\begin{figure*}
    \centering
    \includegraphics[width=\textwidth]{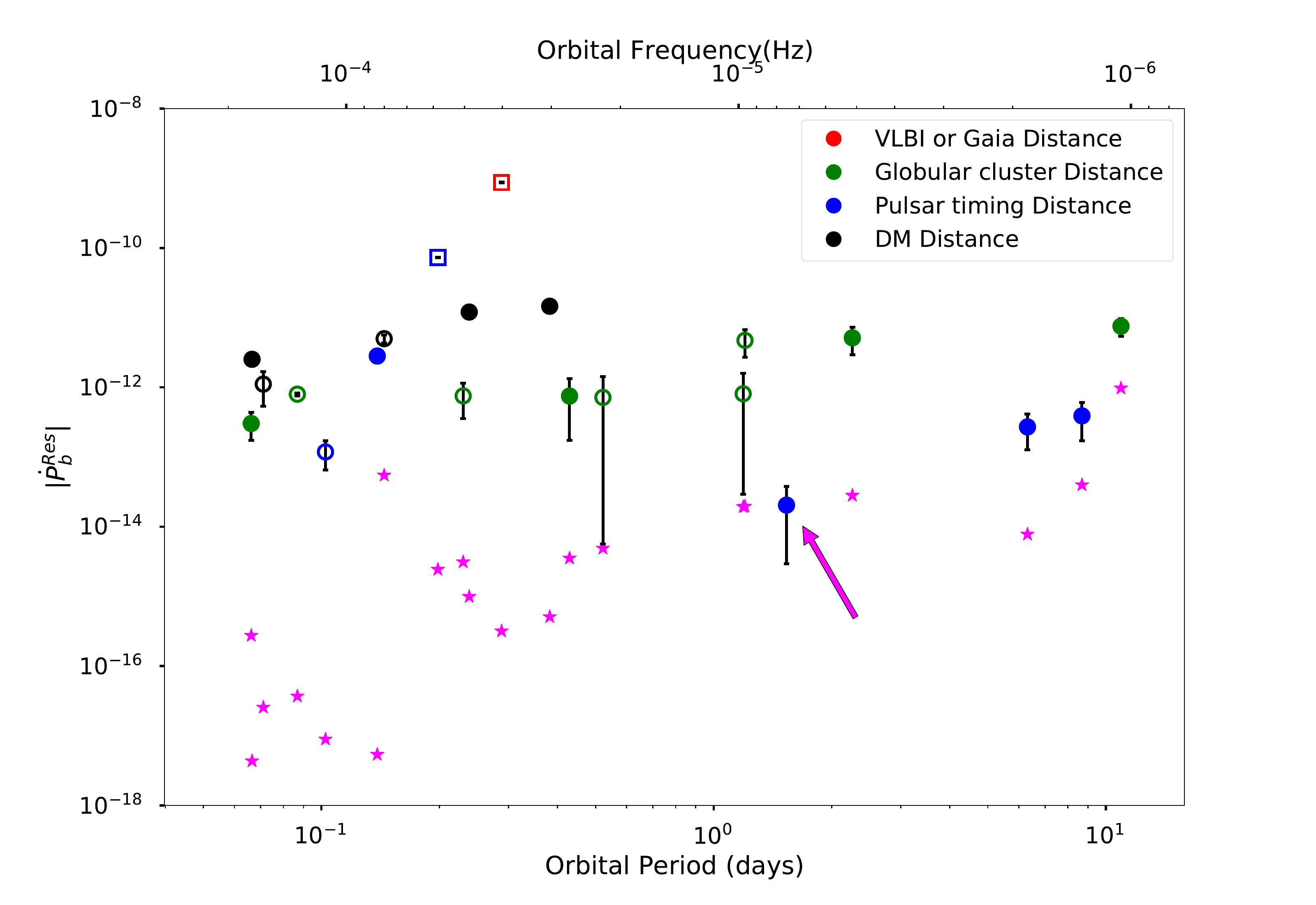}
    \caption{Same as Figure \ref{pbdotfig}, but excluding the fitting line. The magenta star-like points represent the upper limits of the $\dot{P}_b^{\dot{m}}$ for every system with $\dot{P}_b^{Res}$ above a 1 $\sigma$ significance, except for one system with a zero DM uncertainty, which is indicated by the magenta arrow.\label{pbdotdmfig}}
\end{figure*}

For a rough estimation, we assumed that the $1\sigma$ measurement uncertainty of the DM is the upper limit of the amplitude of the DM modulation. Then $m=1.6M_{\odot}$, $\frac{\sin i}{i}=1$, $X=2$, and $v_0=0.1c$ were used for the estimation of the upper limit of the $\dot{m}_c$. For every system, the uncertainties of DM measurement was taken from ATNF \citep{manchester2005australia}, and $a$ was obtained by $a=(Gm/\Omega^2)^{\frac{1}{3}}$. Then the maximum $\dot{P}_b^{\dot{m}}$ was estimated by using equation \ref{mdotwinds} and \ref{mdotofwinds}, as shown in Figure \ref{pbdotdmfig}. One can see that the maximum $\dot{P}_b^{\dot{m}}$ for almost all systems is much smaller than the $\dot{P}_b^{Res}$. So the wind-driven mass loss does not significantly  affect the observational $\dot{P}_b^{Obs}$. 

For the situation in which the pulsar accretes wind from the companion, the second term in equation \ref{mdotwindsacc} should be considered, which is inversely proportional to the companion mass $m_c$. In the other situation in which half of the wind is accreted by the pulsar ($\dot{m}_p=-0.5\dot{m}_c$) with $m_c=0.08 M_{\odot}$ and $m_p=1.4 M_{\odot}$, the $\dot{P}_b^{\dot{m}}$ is about 26 times larger than that in the situation without accretion, which cannot affect the observational $\dot{P}_b^{Obs}$ significantly.

\subsection{Possible origins of the correlation}\label{section discussion correlation}
In almost all tidal models, the tidal response $F(\omega)$ has a power-law relation with the tidal frequency $\omega$ \citep{ogilvie2014tidal}. Whereas, our result predicts a power-law relation between the tidal response $F(\omega)$ and orbital frequency $\Omega$, which requires a power-law relationship between the tidal frequency $\omega$ and the orbital frequency $\Omega$, such a correlation implies a relation between the tidal frequency $\omega$ and the orbital period, which probably stems from the epoch of dynamical co-evolution of spin and orbit.

For typical binaries containing a pulsar with a mass of $m_p=1.48\,M_{\odot}$ and  a WD companion with a mass of $m_c=0.2\,M_{\odot}$, the $\dot{P}_b^{Tide}$ can be equivalent to the $\dot{P}_b^{GW}$ originating in GW radiation at a critical period of $P_c=0.077\,\text{day}$, where the $\dot{P}_b^{Tide}$ was calculated by the power law $K=\alpha P_b^{\beta}$ with the fitting result of the first line in the Table \ref{tab:fitting}. For binaries of orbital period $P_b>P_c$, the tidal interaction determines the change in the orbital period; and for binaries with an orbital period of $P_b<P_c$, the gravitational radiation is  dominant.
 
Consequently, tidal dissipation is responsible for the variation in the orbital period of the binaries with an orbital period of $P_b\approx 1 \,\text{day}$. Even in such a case, the tidal-induced orbital change is only $\dot{P}_b=2\times 10^{-12}ss^{-1}$, which is so small that it takes too long for a binary pulsar to evolve from a long orbital period to a short one. Therefore, the scenario of co-evolution can be ruled out.

Such a correlation between the tidal and orbital frequency can originate in the early formation of the pulsar binary, that is, the recycled stage of a millisecond pulsar binary. Interestingly, all systems of pulsar binaries with a WD companion are millisecond pulsar binaries, which are supposed to have experienced a spin-up stage by accelerating the giant companion, the progenitor of the WD. The accretion of NS through mass transfer results in a wider orbit and reduced envelope of the giant star. As the acceleration comes to a halt, the relics of the giant become a naked core, which is the WD companion.

Theoretically, the property of a remnant WD and orbit can be determined by the initial binary status before the recycled stage. Therefore, a correlation between the physical property of a companion and orbital period in a remnant binary is reasonable. For example, Rappaport reported that the mass of a WD is correlated with the orbital period \citep{rappaport1995relation}. Thus the spin of the companion may correlate with the orbital period.

The progenitor of a companion is usually thought to be synchronized with the orbit during the accretion phase. However, the WD rotates faster than its progenitor giant. Bassa considered a possibility that the envelope had contracted to form the WD after mass transfer \citep{bassa2006masses}. The conservation of angular momentum increases spin and gives $\frac{\omega}{\Omega_b}=\frac{M_{env}R_L^2}{M_{WD}R_{WD}^2}$, where $M_{env}$ and $R_L$ is the mass of the envelope and radius of the Roche-lobe, respectively. This scenario leads to the correlation between companion spin and the orbital period. However, it can only align the spin and the orbit, as well as produce a positive $\dot{P}^{Tide}_{b}$, whereas the negative $\dot{P}^{Tide}_{b}$ corresponding to the opposite spin and orbital direction cannot be explained in this situation.

The negative $\dot{P}^{Tide}_{b}$ represents the negative tidal frequency, which means the spin frequency is either smaller than the orbital frequency in the same direction or an arbitrary value in the opposite direction. In almost all WD-MSP binaries, the spin is larger than the orbital frequency in the dynamical tidal model, so the negative $\dot{P}^{Tide}_{b}$ represents the fast rotation of a companion in the opposite direction as the orbital motion. If the progenitor of a companion has not been synchronized with the orbit and if it has an opposite rotational direction as the orbit, the WD can rotate rapidly in the opposite direction as the orbit.

The tidal dissipation can also affect the evolution of the spin-orbit misalignment angle $\theta$. Lai found that the dumping of the inertial waves in the convective envelope can produce an extra tidal torque to change the spin-orbit inclination misalignment angle $\theta$ \citep{lai2012tidal}, which evolves toward $0$ when $\cos\theta< 0$ and $\pi$ when $\cos\theta> 0$ if the orbital angular momentum is much larger than the spin angular momentum. In the case that the direction of the companion spin was isotropic in the early stage of the binary and that it has been fully evolved in the recycled stage, then the spin-orbit misalignment angle $\theta$ can be either $0$ or $\pi$, both of which having the same likelihood; this corresponds the distribution of the half negative and half positive $\dot{P}_b^{Tide}$, which is consistent with our expectation. 

Now the fast rotation of the companion in both aligned and misaligned situations can be produced, but the origin of the correlation between the companion spin and orbital period remains unknown. Even so, the spin evolution of the companion at the recycled stage is the most probable way to produce such a relation. If the specific relation between the companion spin and orbital period through the recycled process can be understood by other means, the tidal model of WD will be further constrained.
\section{Conclusions}
\label{section conclusion}
There usually are two situations considered in the literature: one is that the companion star is already synchronized with orbit, and the other is that the tidal dissipation on the WD is inefficient. Thus the existence of tidal dissipation in pulsar binaries with the WD companion has long been neglected. Fuller's study on the dynamical tide of the WD \citep{fuller2012dynamical} predicts a more efficient tidal dissipation than viscosity turbulent in Zahn's equilibrium tidal model \citep{zahn1989tidal}. On the other hand, the time interval between the recycled stage to the present day is difficult to estimate, and it is not possible to guarantee that the companion has been synchronized with the orbit. Therefore, it is too early to exclude the effect of tidal dissipation from the timing of these pulsar binaries.

After counting out the three normal effects, the residual term $\dot{P}_b^{Res}$ displays signs of being half positive and half negative, which is compatible with the scenario of the tidal effect. We further found that the dynamical tide on the WD can be efficient enough to account for the residual term $\dot{P}_b^{Res}$ of all PSR-WD binaries. For the PSR-MS binary, the Love number of the MS companion of \object{PSR J1227-4853} can be constrained as $k_2=0.177^{+0.098}_{-0.058}$, which is a reasonable value for the MS. This supports the possibility of tidal dissipation in the PSR-MS system. Therefore, tidal dissipation should not be ignored in the orbital decay $\dot{P}_b$.

Furthermore, a weak correlation between the tidal parameter and orbital period exhibits a potential relation between the tidal frequency and orbital frequency. As all of the systems are millisecond pulsar binaries, this suggests that such a correlation likely originates from the recycled stage of millisecond pulsar binaries.
 
Investigation of the tidal dissipation on pulsar binaries is of importance in three aspects. First, it indicates that tidal dissipation also contributes to the first derivative of the orbital period of pulsar binaries, so that fundamental physics such as gravitational radiation of a binary pulsar should be extracted after counting it out. Second, it shows that the interior of the companion stars can be investigated through the precise pulsar timing of these pulsar binaries, which provides a useful complement to previous methods, such as spectroscopy,  light curves, eclipse, and asteroseismology. Third, the derived correlation between the tidal and orbital frequency suggests that tidal dissipation may play an important role in the recycled stage of a millisecond pulsar binary.

\begin{acknowledgements}
    This work is supported by the Ministry of Science and Technology of the People’s Republic of China (2020SKA0120300).
\end{acknowledgements}

\bibliography{ref} 
\bibliographystyle{aa}

\begin{appendix}
\section{Collected Data}
\label{data table}

\begin{table*}[htp]
    \centering
    \caption{Measurement of sine of inclination $\sin (i)$, companion mass $m_c$, and mass ratio $m_p/m_c$.\label{shaprio}}
    \begin{tabular}{ccccc}
    Name&$\sin (i)$&$m_c$&$m_p/m_c$&Reference\\
    \hline
    &&$M_{\odot}$&&\\
    \hline
        J0348+0432$^*$&*&0.173(3)&11.70(13)&\citet{cognard2017massive}\\
        J0437-4715$^*$&0.6748(5)&0.224(7)&*&\citet{reardon2015timing}\\
        J1012+5307$^*$&*&0.16(2)&10.0(7)&\citet{van2004optical}\\
        J1614-2230$^*$&0.999903(3)&0.493(3)&*&\citet{arzoumanian2018nanograv}\\
        J1909-3744$^*$&0.9981(1)&0.208(2)&*&\citet{arzoumanian2018nanograv}\\
        J2222-0137&0.9966(3)&1.293(25)&*&\citet{cognard2017massive}\\
        J0024-7204S&*&0.17(3)&*&\citet{cadelano2015optical}\\
        J1024-7204U&*&0.171(3)&*&\citet{cadelano2015optical}\\
        J0024-7204Y&*&0.17(2)&*&\citet{cadelano2015optical}\\
        J1227-4853&*&*&5.15(8)&\citet{de2014unveiling}\\
        J1738+0333&*&*&8.1(2)&\citet{freire2012relativistic}\\
    \hline
    \end{tabular}
\end{table*}

\begin{table*}[htp]
    \centering
    \caption{Parameter of globular clusters where binaries are located and derived by Baumgardt \citep{baumgardt2018mean}\label{gc}}
    \begin{tabular}{ccccc}
        Name of globular clusters&$\log \rho_c$&$r_c$&Distance&Pulsar inside\\
        \hline
        &$M_{\odot}/pc^3$&pc&kpc&\\
        \hline
        &&&&J0024-7204Y,
        J0024-7204X,
        J0024-7204U\\
        NGC 104&5.12&0.51&$4.43\pm0.04$&
        J0024-7204S
        J0024-7204R
        B0021-72E\\
        &&&&
        J0024-7204Q,
        B0021-72I\\
        \hline
        NGC 5904&3.67&1.19&$7.57\pm 0.13$&J1518+0204C\\
        \hline
        NGC 6266&5.21&0.36&$6.41\pm 0.12$&J1701-3006B\\
        \hline
        NGA6544&6.28&0.16&$2.6\pm 0.27$&1807-2459A\\
        \hline
    \end{tabular}
\end{table*} 

\begin{sidewaystable}[p]
\label{tab:data}
\centering
\caption{ Observational and derived values of our samples. The distance was measured by the globular cluster for Group A, pulsar timing for Group B, and the DM for Group C; also two systems with a distance measured by the globular cluster that do not have a proper motion measurement also belong to Group C. Other systems were not fitted because they do not have  $\dot{P}^{Res}$ within a 1 $\sigma$ significance. Three asterisks represent the systems having $\dot{P}^{Res}$ within a 3 $\sigma$ significance; two asterisks represent the systems having $\dot{P}^{Res}$ within a 2 $\sigma$ significance; the asterisk represents the systems having $\dot{P}^{Res}$ within a 1 $\sigma$ significance.}
\footnotesize
\setlength\tabcolsep{3pt}
\begin{tabular}{ccccccccccccc}
Name&Mass function&Comp&$P_b$&$\dot{P}_b^{obs}$&$\mu_{\alpha}$&$\mu_{\delta}$&Distance&$\dot{P}_b^{Shk}$&$\dot{P}_b^{Acc}$&$\dot{P}_b^{GW}$&$\dot{P}_b^{Res}$&References\\
\hline
&($M_{\odot}$)&&(days)&&(mas $\text{yr}^{-1}$)&(mas $\text{yr}^{-1}$)&(kpc)&&&&&\\
\hline
\multicolumn{13}{c}{Group A}\\
\hline
J0024-7204X$^{***}$&$1.52\times 10^{-2}$&WD&10.92 &      $6(2) \times 10^{-12}$ &     $5.8(1)$ &     $-3.3(2)$ &   $4.43(4)$ &   $4.5(2) \times 10^{-13}$ &  $-2.0^{+0.6}_{-0.6}\times 10^{-12}$ &       $-2.2^{+0.5}_{-1}\times 10^{-16}$ &            $8^{+2}_{-2}\times 10^{-12}$&2,3,5,25\\
B0021-72E$^{**}$&$1.64\times 10^{-3}$&WD&2.257 &    $4.8(2) \times 10^{-12}$ &     $6.15(3)$ &    $-2.35(6)$ &   $4.43(4)$ &   $9.1(1) \times 10^{-14}$ &        $-4^{+9}_{-9}\times 10^{-13}$ &     $-1.4^{+0.3}_{-0.7}\times 10^{-15}$ &      $5.1^{+0.9}_{-0.9}\times 10^{-12}$&2,3,25\\
J0024-7204R$^{**}$&$9.09\times 10^{-6}$&-&0.06623 &    $1.9(4) \times 10^{-13}$ &     $4.8(1)$ &     $-3.3(2)$ &   $4.43(4)$&  $2.09(9) \times 10^{-15}$ &        $-1^{+8}_{-8}\times 10^{-14}$ &           $-9^{+2}_{-4}\times 10^{-14}$ &      $2.0^{+0.9}_{-0.9}\times 10^{-13}$ &2,3,25\\
J0024-7204S$^{**}$&$3.35\times 10^{-4}$&WD&1.202 &   $-4.9(4) \times 10^{-12}$ &     $4.5(1)$ &    $-2.5(1)$ &   $4.43(4)$ &   $3.0(1) \times 10^{-14}$ &      $-2^{+14}_{-15}\times 10^{-13}$ &       $-2.3^{+0.5}_{-1}\times 10^{-15}$ &           $-5^{+1}_{-1}\times 10^{-12}$&2,3,4,25\\
J1518+0204C$^{**}$&$2.68\times 10^{-5}$&-&0.08683 &   $-9.14(23) \times 10^{-13}$ &      $4.67(14)$ &     $-8.24(36)$ &    $7.57(13)$ &$1.24(8)\times 10^{-14}$ &        $-4^{+1}_{-1}\times 10^{-14}$ &           $-8^{+2}_{-4}\times 10^{-14}$ &     $-8.8^{+0.3}_{-0.3}\times 10^{-13}$ &16,25\\
B0021-72I$^{*}$&$1.16\times 10^{-6}$&-&0.2298 &     $-8(2) \times 10^{-13}$ &      $5.0(2)$ &     $-2.1(2)$ &   $4.43(4)$ &   $6.3(4) \times 10^{-15}$ &      $-4^{+23}_{-22}\times 10^{-14}$ &           $-5^{+1}_{-3}\times 10^{-15}$ &           $-8^{+3}_{-3}\times 10^{-13}$&2,3,4,25\\
J0024-7204Q$^{*}$&$2.37\times 10^{-3}$&WD&1.189 &    $-1.0(2) \times 10^{-12}$ &     $5.2(1)$ &    $-2.6(1)$ &   $4.43(4)$&   $3.7(1) \times 10^{-14}$ &        $-2^{+2}_{-2}\times 10^{-13}$ &           $-5^{+1}_{-2}\times 10^{-15}$ &           $-8^{+3}_{-3}\times 10^{-13}$ &2,3,4,25\\
J0024-7204U$^{*}$&$8.53\times 10^{-4}$&WD&0.4291 &    $6.6(5) \times 10^{-13}$ &      $4.6(2)$ &    $-3.8(1)$ &   $4.43(4)$&  $1.42(8) \times 10^{-14}$ &      $-8^{+24}_{-24}\times 10^{-14}$ &     $-1.8^{+0.4}_{-0.9}\times 10^{-14}$ &            $7^{+2}_{-2}\times 10^{-13}$ &2,3,4,25\\
J0024-7204Y$^{*}$&$1.18\times 10^{-3}$&WD&0.5219 &   $-8.2(7) \times 10^{-13}$ &     $4.4(1)$ &    $-3.4(1)$ &   $4.43(4)$&   $1.5(5) \times 10^{-14}$ &     $-10^{+29}_{-29}\times 10^{-14}$ &     $-1.4^{+0.3}_{-0.8}\times 10^{-14}$ &           $-7^{+3}_{-3}\times 10^{-13}$ &2,3,4,25\\
\hline
\multicolumn{13}{c}{Group B}\\
\hline
J0023+0923$^{***}$&$2.00\times 10^{-6}$&-&  0.1388 &    $2.8(2) \times 10^{-12}$ &    $-12.6(2)$ &     $-5.8(3)$ &    $1.1(2)$&     $6(1) \times 10^{-15}$ &  $-2.2^{+0.5}_{-0.5}\times 10^{-15}$ &     $-1.5^{+0.3}_{-0.8}\times 10^{-14}$ &      $2.8^{+0.2}_{-0.2}\times 10^{-12}$&1\\
J0348+0432$^{**}$&$2.87\times 10^{-4}$&WD&0.1024 &   $-2.73(45) \times 10^{-13}$ &      $4.04(16)$ &      $3.5(6)$ &    $2.1(2)$ &   $1.3(2) \times 10^{-15}$ &         $6^{+2}_{-2}\times 10^{-16}$ &     $-1.6^{+0.3}_{-0.3}\times 10^{-13}$ &     $-2.8^{+0.4}_{-0.4}\times 10^{-13}$&6\\
J1603-7202$^{*}$&$8.79\times 10^{-3}$&WD&6.309 &      $3.1(15) \times 10^{-13}$ &    $-2.46(4)$ &    $-7.33(5)$ &      $0.9(7)$ &     $7(5) \times 10^{-14}$ &        $-2^{+2}_{-4}\times 10^{-14}$ &       $-4.6^{+1.0}_{-3}\times 10^{-16}$ &            $3^{+1}_{-1}\times 10^{-13}$&7,10\\
J1614-2230$^{*}$&$2.05\times 10^{-2}$&WD&8.687 &    $1.7(2) \times 10^{-12}$ &     $3.8(10)$ &    $-32.5(7)$ &    $0.67(5)$ &   $1.3(1) \times 10^{-12}$ &   $2.6^{+0.7}_{-0.6}\times 10^{-15}$ &   $-4.2^{+0.05}_{-0.05}\times 10^{-16}$ &            $4^{+2}_{-2}\times 10^{-13}$&1,19\\
J1909-3744$^{*}$&$3.12\times 10^{-3}$&WD&1.533 &   $5.02(5) \times 10^{-13}$ &   $-9.516(4)$ &  $-35.77(1)$ &   $1.09(4)$ &   $4.8(2) \times 10^{-13}$ &   $3.5^{+0.3}_{-0.3}\times 10^{-15}$ &  $-2.77^{+0.05}_{-0.05}\times 10^{-15}$ &            $2^{+2}_{-2}\times 10^{-14}$&1,7,10\\
\hline
\multicolumn{13}{c}{Group C}\\
\hline
J1701-3006B$^{***}$&$8.83\times 10^{-4}$&WD&0.1445 &   $-5.12(62) \times 10^{-12}$ &        -      &        -      &    $6.4(1)$ &       - &     $6^{+235}_{-239}\times 10^{-15}$ &     $-1.1^{+0.2}_{-0.6}\times 10^{-13}$ &     $-5.1^{+0.6}_{-0.6}\times 10^{-12}$&3,15,25\\
J1957+2516$^{***}$&$4.31\times 10^{-4}$&WD&0.2381 &   $1.2(2) \times 10^{-11}$ &     $-5.68(113)$ &    $-8.86(147)$ &      $2.66$ &   $1.5(4) \times 10^{-14}$ &  $-1.2^{+0.4}_{-0.6}\times 10^{-14}$ &       $-3.8^{+0.8}_{-2}\times 10^{-14}$ &      $1.2^{+0.2}_{-0.2}\times 10^{-11}$&3,14,20,23\\
B1957+20$^{***}$&$5.00\times 10^{-6}$&-&0.3820 &   $1.47(8) \times 10^{-11}$ &     $-16.0(5)$ &    $-25.8(6)$ &      $1.73$&   $1.3(2) \times 10^{-13}$ &        $-7^{+2}_{-3}\times 10^{-15}$ &       $-3.8^{+0.8}_{-2}\times 10^{-15}$ &   $1.46^{+0.07}_{-0.08}\times 10^{-11}$ &3,14\\
J0636+5129$^{***}$&$1.76\times 10^{-7}$&-&0.06655 &    $2.5(3) \times 10^{-12}$ &      $3.5(2)$ &     $-2.3(2)$ &      $0.21$&     $5(1) \times 10^{-17}$ &         $3^{+2}_{-1}\times 10^{-18}$ &       $-2.3^{+0.5}_{-1}\times 10^{-14}$ &      $2.5^{+0.3}_{-0.3}\times 10^{-12}$ &1,3\\
J1807-2459A$^{*}$&$3.88\times 10^{-7}$&-&0.07109 &  $-1.14(6) \times 10^{-12}$ &        -      &        -      &    $2.6(3)$&       - &     $1^{+568}_{-566}\times 10^{-15}$ &       $-2.7^{+0.6}_{-1}\times 10^{-14}$ &     $-1.1^{+0.6}_{-0.6}\times 10^{-12}$ &3,15,25\\
\hline
\multicolumn{13}{c}{No Fitting}\\
\hline
J0437-4715&$1.24\times 10^{-3}$&WD& 5.741 &  $3.728(6) \times 10^{-12}$ &  $121.438(2)$ &  $-71.475(2)$ &   $0.157(2)$ &  $3.76(5) \times 10^{-12}$ &  $-3.7^{+0.1}_{-0.1}\times 10^{-15}$ &     $-3.2^{+0.2}_{-0.2}\times 10^{-16}$ &           $-2^{+5}_{-4}\times 10^{-14}$&7\\
J0751+1807&$9.67\times 10^{-4}$&WD&0.2631 &   $-3.50(25) \times 10^{-14}$ &    $-2.73(5)$ &    $-13.4(3)$ &    $1.07(24)$&   $1.1(2) \times 10^{-14}$ &         $6^{+5}_{-3}\times 10^{-16}$ &       $-4.2^{+0.9}_{-2}\times 10^{-14}$ &     $-4.7^{+0.4}_{-0.4}\times 10^{-14}$  &8,9\\ 
J1012+5307&$5.78\times 10^{-4}$&WD&0.6047 &      $8.1(16) \times 10^{-14}$ &     $2.66(3)$ &    $-25.5(4)$ &    $1.15(24)$ &    $10(2) \times 10^{-14}$ &  $-4.6^{+0.7}_{-0.5}\times 10^{-15}$ &           $-9^{+2}_{-2}\times 10^{-15}$ &           $-1^{+2}_{-2}\times 10^{-14}$&1,8,10\\
J1023+0038&$1.11\times 10^{-3}$&MS&0.1981 &  $-7.32(6) \times 10^{-11}$ &     $4.76(3)$ &   $-17.34(4)$ &   $1.368(42)$&  $1.84(5) \times 10^{-14}$ &  $-3.3^{+0.1}_{-0.1}\times 10^{-15}$ &           $-7^{+2}_{-4}\times 10^{-14}$ &  $-7.32^{+0.06}_{-0.06}\times 10^{-11}$ &11,12,13\\
J1227-4853&$3.87\times 10^{-3}$&MS&0.2879 &  $-8.7(1) \times 10^{-10}$ &    $-18.726(207)$ &      $7.338(116)$ &    $1.6(4)$&   $3.9(9) \times 10^{-14}$ &        $-6^{+2}_{-3}\times 10^{-15}$ &           $-6^{+1}_{-1}\times 10^{-14}$ &   $-8.7^{+0.09}_{-0.09}\times 10^{-10}$ &22,23,24\\
J1738+0333&$3.46\times 10^{-4}$&WD&0.3548 &   $-1.70(31) \times 10^{-14}$ &    $7.037(5)$ &     $5.07(1)$ &  $1.47(10)$ &   $8.2(5) \times 10^{-15}$ &        $-8^{+1}_{-2}\times 10^{-16}$ &     $-2.8^{+0.6}_{-0.6}\times 10^{-14}$ &  $-2.45^{+0.05}_{-0.05}\times 10^{-14}$&10,21\\
J2222-0137&$2.29\times 10^{-1}$&WD&2.446 &    $2.0(9) \times 10^{-13}$ &    $44.73(2)$ &    $-5.68(6)$ &   $26.7(1)$ &  $2.79(1) \times 10^{-13}$ &  $-4.6^{+0.1}_{-0.1}\times 10^{-15}$ &     $-7.7^{+0.3}_{-0.3}\times 10^{-15}$ &           $-7^{+8}_{-8}\times 10^{-14}$&17,18\\

\hline
\end{tabular}
\tablebib{
(1)~\citet{arzoumanian2018nanograv}; (2) \citet{freire2017long}; (3) \citet{manchester2005australia}; (4) \citet{cadelano2015optical};
(5) \citet{ridolfi2016long}; (6) \citet{antoniadis2013massive}; (7) \citet{reardon2015timing}; (8) \citet{desvignes2016high};
(9) \citet{bassa2006ultra}; (10) \citet{van2004optical}; (11) \citet{archibald2013long}; (12) \citet{deller2012parallax}; (13) \citet{bogdanov2015coordinated}; (14) \citet{arzoumanian1993orbital}; (15) \citet{lynch2012timing}; (16) \citet{pallanca2014radio}; (17) \citet{cognard2017massive}; (18) \citet{deller2013vlbi}; (19) \citet{bhalerao2011white}; (20) \citet{stovall2016timing}; (21) \citet{freire2012relativistic}; (22) \citet{roy2015discovery}; (23) \citet{gaia2018vizier}; (24) \citet{de2014unveiling}; (25) \citet{baumgardt2018mean}.
}
\end{sidewaystable}

\end{appendix}

\end{document}